\documentclass[conference]{IEEEtran}
\IEEEoverridecommandlockouts
\usepackage[utf8]{inputenc}
\usepackage[T1]{fontenc}
\usepackage{cite}
\usepackage{amsmath,amssymb,amsfonts}
\usepackage{graphicx}
\usepackage{textcomp}
\usepackage{xcolor}
\usepackage{booktabs}
\usepackage{array}
\usepackage{enumitem}
\usepackage{url}
\usepackage[hidelinks]{hyperref}

\begin{document}

\title{Toward a Full-Stack Framework for Industrial Augmented Reality:\\
Benefits, Risks, and Design Considerations for Dependable Deployment in Manufacturing}

\author{%
  \IEEEauthorblockN{Narges Chinichian and Maximilian Anton Palm}\\
  \IEEEauthorblockA{\textit{MixedForm LTD, Dublin, Ireland} \\
    narges.chinichian@mixedform.com, \\
    maximilian.palm@mixedform.com}
}

\maketitle

\begin{abstract}
Industrial Augmented Reality (AR) has progressed from laboratory demonstrations to operational pilots across design, training, assembly, maintenance and quality assurance, yet broad, dependable deployment in manufacturing remains the exception. We synthesise existing evidence into a \emph{full-stack} deployment framework structured along six distinct but coupled decision axes: (i) value and benefits, (ii) technical and integration constraints, (iii)~human factors and safety, (iv) organisational and economic considerations, (v) data, security and privacy, and (vi) governance, ethics and long-term risk. Within each axis we separate \emph{(a)benefits}, \emph{(b)failure modes} and \emph{(c)design considerations}, and cross-link them through a deployment checklist that engineering managers and vendors can apply when scoping projects. The contribution is conceptual and practice-oriented: a synthesis grounded in the literature and public deployment reports. We mark where the evidence base is mature (e.g.\ assembly task time, training efficacy), emerging (e.g.\ cognitive workload trade-offs, cobot safety zones), or speculative (e.g.\ metaverse-scale governance), and identify open questions whose resolution conditions the transition from demos to dependable infrastructure.
\end{abstract}


% =============================================================================
\section{Introduction}
\label{sec:intro}

Augmented Reality (AR), as defined by Azuma \cite{azuma1997survey} and situated on Milgram and Kishino's reality--virtuality continuum \cite{milgram1994taxonomy}, has been promoted for over two decades as a key human interface for digitised manufacturing. Systematic reviews document a steady accumulation of applications in design, assembly, maintenance, quality assurance, training and factory planning \cite{egger2020ar,fang2023hmdreview,runji2022maintenance,solomashenko2025assisted}. Industry programmes increasingly position AR as the human-facing layer of the \emph{industrial metaverse}, in which digital twins, cyber-physical systems and Extended Reality converge to support Industry~4.0/5.0 operations \cite{guo2024industrialmetaverse,kour2025metaverseindustrial,yang2025mfgverse}.

Despite this momentum, deployments at scale remain rare. Adoption studies consistently report that the gap between feasibility and dependable industrial use is dominated by software quality, integration friction, mental workload, organisational readiness and acceptance, rather than by display hardware alone \cite{quandt2021useracceptance,roberto2024commercialar,machac2025arampaper}. In parallel, Extended Reality has surfaced a new generation of risks (biometric inference, gaze fingerprinting, attentional tunnelling, and large infrastructure footprints) that the industrial deployment literature has only begun to internalise \cite{anjum2025xrhealthsok,kundu2025privatexr,nleya2024industrialmetaverse}.

This paper argues that moving industrial AR from \emph{demos} to \emph{dependable infrastructure} requires a deliberately multi-axis treatment in which benefits, failure modes and design considerations are addressed jointly rather than in isolated technology or human-factors silos. We synthesise the evidence into a six-axis framework (Section~\ref{sec:framework}), make each axis operational through a deployment checklist (Section~\ref{sec:checklist}) and conclude with open research questions (Section~\ref{sec:open}).

\paragraph*{Scope and method.}
This is a conceptual synthesis. Where evidence is uneven we label claims as \emph{established consensus}, \emph{emerging research} or \emph{speculative}. The scope is discrete and process manufacturing; surgical and automotive-HUD studies are cited only where they ground transferable findings.
% =============================================================================

\begin{figure*}[t]
  \centering
  \includegraphics[width=0.9\textwidth]{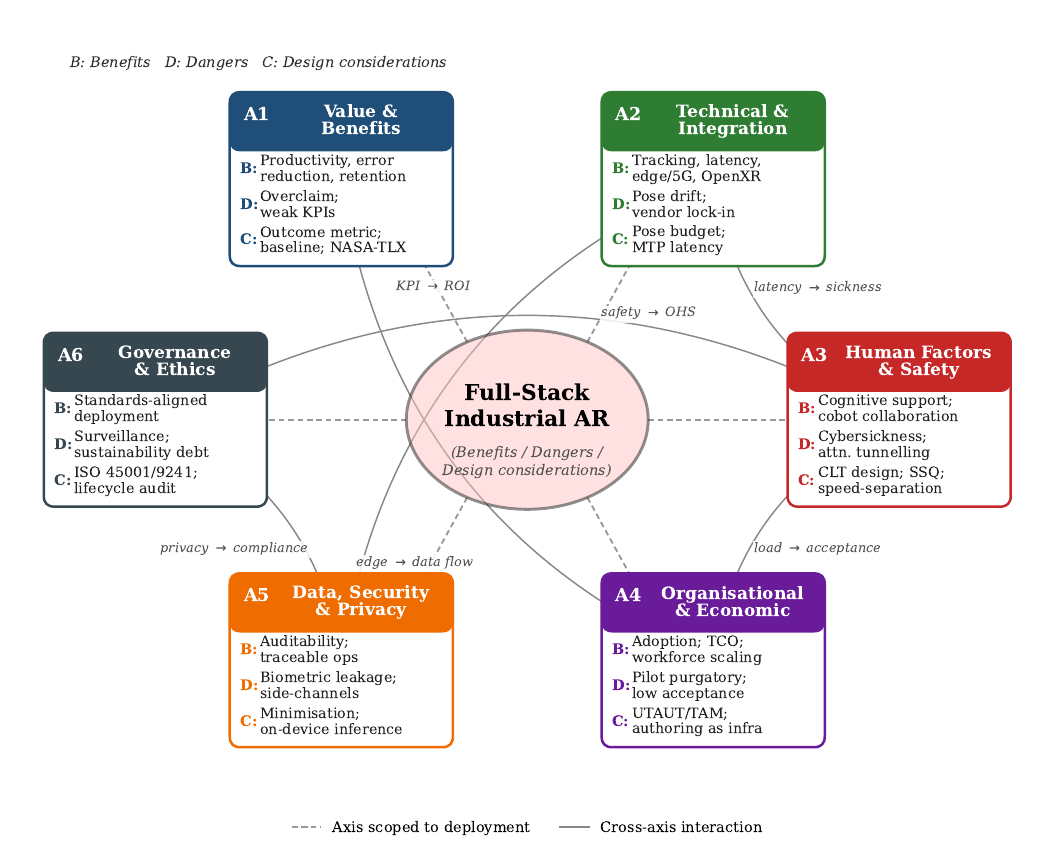}
  \caption{Six-axis framework for full-stack industrial AR. Axes A1--A6 (value, technical/integration, human factors \& safety, organisational/economic, data \& privacy, governance \& ethics) are each examined through three lenses, Benefits (B), Dangers (D) and Design considerations (C). Dashed radial connectors indicate that each axis is scoped to the deployment; solid arrows highlight representative cross-axis couplings (e.g.\ latency$\rightarrow$sickness, privacy$\rightarrow$compliance). Abbreviations: AR = Augmented Reality, CLT = Cognitive Load Theory, ISO = International Organization for Standardization, KPI = Key Performance Indicator, MTP = Motion-to-Photon, OHS = Occupational Health and Safety, ROI = Return on Investment, SSQ = Simulator Sickness Questionnaire, TCO = Total Cost of Ownership, UTAUT/TAM = Unified Theory of Acceptance and Use of Technology/Technology Acceptance Model}
  \label{fig:six_axis}
\end{figure*}
% =============================================================================
\section{Background and Related Frameworks}
\label{sec:background}

Existing reviews tend to organise industrial AR by \emph{application area}(design \cite{baroroh2020smartmanuf}, training \cite{howard2023metaanalysis}, maintenance \cite{runji2022maintenance,buettner2022vrarmaint}, assembly \cite{yang2019stages,sahoo2025laser}, and quality \cite{ho2022quality,seeliger2023qualityinspection}) or by \emph{enabling technology}, including head-mounted displays Head-Mounted Displays, spatial AR and mobile/handheld AR \cite{fang2023hmdreview,alves2019sparvshandheld,solomashenko2025assisted}. A smaller line proposes user-acceptance models tailored to industrial AR, typically extensions of Unified Theory of Acceptance and Use of Technology (UTAUT) or Technology Acceptance Model (TAM) \cite{quandt2021useracceptance,marto2023aram,machac2025arampaper}. Industrial-metaverse reviews adopt a higher-level lens, foregrounding cyber-physical convergence and architecture but providing limited operational guidance \cite{guo2024industrialmetaverse,nleya2024industrialmetaverse,kour2025metaverseindustrial}.

What is missing, and what we attempt here, is a \emph{deployment-oriented} synthesis that explicitly couples benefits, dangers and design considerations across technical, human, organisational, security and governance layers, so that decisions taken in one layer can be reasoned about in terms of their consequences in the others.

% =============================================================================
\section{A Full-Stack Framework for Industrial Augmented Reality}
\label{sec:framework}

We propose six axes (A1--A6). Each is treated under three lenses: \textbf{Benefits} (value levers), \textbf{Dangers} (failure modes and risks) and \textbf{Design considerations} (concrete factors for specification or selection). The axes are \emph{distinct but coupled} decision axes in the sense of separation-of-concerns: they each pose decisions that cannot be reduced to the others, but they interact strongly in operation. Section~\ref{sec:checklist} makes the recurring couplings explicit.

\paragraph*{Provenance of the axes.}
The framework is a synthesis, not a discovery. A1--A3 are the \emph{consensus axes} that recur across industrial-Augmented Reality (AR) reviews and were found relevant in our own implementation of AR in an industrial setting: value/benefits, technical/integration constraints, and human factors/safety appear in every systematic review surveyed \cite{egger2020ar,fang2023hmdreview,runji2022maintenance,solomashenko2025assisted,baroroh2020smartmanuf}. A4 (organisational/economic) appears in most adoption-focused work but is treated unevenly \cite{quandt2021useracceptance,roberto2024commercialar,machac2025arampaper}. A5 (data/security/privacy) and A6 (governance/ethics) are the contribution: they are well developed in adjacent literatures (Extended Reality-security systematisations \cite{anjum2025xrhealthsok,elhajj2024xrsec,kundu2025privatexr} and Occupational Health and Safety/ergonomics standards \cite{antonov2017iso45001,mugeere2023iso9241})but are typically under-weighted in industrial-AR reviews, despite first-order effects on deployment. The framework's claim is therefore that A5 and A6 must be treated as peers of A1--A3, not as downstream legal afterthoughts.

\subsection{A1: Value and Benefits}
\label{sec:a1value}

\textbf{Benefits.}
The strongest evidence concerns assembly and maintenance: AR guidance reduces task time and error rate relative to paper or screen-based instructions across multiple controlled studies and systematic reviews \cite{yang2019stages,alessa2023arcogload,alves2019sparvshandheld,runji2022maintenance,fang2023hmdreview}. Training meta-analyses report consistent improvements in skill performance and learning outcomes broadly, with
overall effect sizes of approximately 0.6 SD that are robust though heterogeneous \cite{howard2023metaanalysis,baashar2022armedical,gong2024arsafetytraining}. AR is numerically most effective for procedural skills (d=0.73) relative to declarative knowledge (d=0.62) \cite{howard2023metaanalysis}. Meta-analytic evidence indicates that AR safety training produces a significant overall positive effect relative to conventional approaches, though no statistically significant difference was found for knowledge acquisition, and evidence on longer-term knowledge retention remains limited \cite{gong2024arsafetytraining}. Quality-assurance benefits are emerging but real \cite{ho2022quality,seeliger2023qualityinspection}.

\textbf{Dangers.}
Benefit estimates are dominated by small-sample lab studies; few report long-term skill retention, line-level productivity, or generalisation beyond the trained task \cite{howard2023metaanalysis,machac2025arampaper}. Knowledge-versus-skill dissociation is documented: AR improves time and confidence more reliably than declarative knowledge \cite{baashar2022armedical}.

\textbf{Design considerations.}
Specify the value lever ex ante (cycle time, defect rate, training cost, expert scarcity) and choose evaluation metrics that match (e.g.\ NASA-Task Load Index plus task time for assembly, retention plus transfer for training \cite{alessa2023arcogload,howard2023metaanalysis}). Avoid privileging novelty-driven engagement metrics over operational ones.

\subsection{A2: Technical and Integration Constraints}
\label{sec:a2technical}

\textbf{Benefits.}
Modern Head-Mounted Displays, optical see-through displays and projector-based spatial AR provide spatially registered overlays that, when stable, support hands-free operation, contextual information delivery and collaboration with cobots and digital twins \cite{fang2023hmdreview,alves2019sparvshandheld,emporio2024xrfactory}. 5G and multi-access edge computing (MEC) reduce on-device load and latency, enabling split rendering and multi-user remote collaboration \cite{sampath2024splitxr,trinh2023drlxr,aktas2025aislicingxr}. Integration with digital twins enables bidirectional information flow between the shop floor and AR overlays \cite{emporio2024xrfactory,xu2024arcam,yang2025mfgverse}.

\textbf{Dangers.}
Tracking is the most consistently reported failure mode in industrial settings: in factory-scale benchmarks of commercial Simultaneous Localization and Mapping (SLAM) stacks (ARKit, ARCore, HoloLens), Feigl et al.\ measured an average position error on the order of $17$~m over a $\sim 1600$~m\textsuperscript{2} area, with a scaling error of roughly $14.4$~cm per metre walked along $120$~m paths between cubicles spaced $\sim 30$~m apart, enough to render world-locked overlays unusable in large, dynamic factory zones \cite{feigl2020localizationlimits,haischt2023automotivetracking}. Network conditions in real plants frequently violate the latency budgets required for stable visual--vestibular fusion; split-rendering architectures can shift but not eliminate this constraint \cite{sampath2024splitxr,trinh2023drlxr}. Interoperability is fragmented: OpenXR has materially reduced platform lock-in but heterogeneity in tracking, input and rendering pipelines persists, and industrial Internet of Things (IoT) platforms remain poorly harmonised \cite{jang2023crossplatformxr,koo2025iiotinterop}. Proprietary firmware and limited bi-directional data exchange are routinely cited as adoption blockers \cite{roberto2024commercialar,machac2025arampaper}.

\textbf{Design considerations.}
Match the tracking modality (marker, model-constrained Simultaneous Localization and Mapping (SLAM), hybrid visual--inertial, projector-based) to the spatial scale, lighting and environmental dynamics of the cell, and budget for calibration drift, including hand-tracking reliability where bare-hand input is part of the workflow \cite{feigl2020localizationlimits,wachter2024spatialar}. Specify motion-to-photon latency targets and verify them under realistic radio conditions, not lab Wi-Fi \cite{sampath2024splitxr}. Require OpenXR support, documented data interfaces (e.g.\ to digital-twin services) and an explicit data-residency model \cite{jang2023crossplatformxr,emporio2024xrfactory}.

\subsection{A3: Human Factors and Safety}
\label{sec:a3human}

\textbf{Benefits.}
Properly designed AR reduces extraneous cognitive load by co-locating information with the task, supporting germane processing and longer-term skill formation \cite{yang2019stages,alessa2023arcogload}. Mixed-reality interfaces communicating cobot intent improve human--robot trust, comfort and efficiency \cite{zhang2025mrhrc,cogurcu2023safetyzones,orsolits2025dynamicsafetyzones}.

\textbf{Dangers.}
Head-Mounted Display use produces measurable cybersickness, oculomotor strain and discomfort, with prevalence and severity strongly dependent on display type, content dynamics and context; effects are amplified for video-passthrough relative to direct vision or optical see-through (driven by depth/scale mismatch and motion-parallax artefacts) and for moving-platform use, and reduced but not eliminated by mitigation techniques such as geometry-aware passthrough rendering \cite{freiwald2018camera,szentirmai2025you,el2025mind,bailenson2024passthrough}. Attentional tunnelling and inattentional blindness are a distinct and underappreciated risk: operators systematically miss non-augmented but safety-critical stimuli, with the effect amplified when overlays are placed centrally and task load is high \cite{wang2021ibarhud,chen2023ibposition,faria2025ibhuds,ye2025arhudsafety}. Ergonomics is a persistent constraint: Head-Mounted Display weight, optical fit and prolonged head posture contribute to musculoskeletal load \cite{okeeffe2024arhmdshipbuilding,fang2023hmdreview}.

\textbf{Design considerations.}
Apply Cognitive Load Theory to overlay design (minimise extraneous load, sustain germane load) and validate with multi-modal measurement (NASA-Task Load Index, eye-tracking, EEG/HRV where ethically appropriate) \cite{alessa2023arcogload,candido2025multimedialearning}. Treat attentional tunnelling as a first-class hazard: preserve peripheral salience, avoid central occlusion of likely hazard locations and test under realistic dual-task conditions \cite{wang2021ibarhud,faria2025ibhuds}. For cobot interaction, use MR-visualised dynamic safety zones aligned with speed-and-separation monitoring, not static fences \cite{cogurcu2023safetyzones,orsolits2025dynamicsafetyzones}. Plan exposure schedules to limit cumulative oculomotor and postural load, and screen for susceptible users \cite{cossio2025cybersickness}.

\subsection{A4: Organisational and Economic Considerations}
\label{sec:a4org}

\textbf{Benefits.}
AR amortises expert knowledge by codifying procedures into reusable overlays and enabling remote assistance, which is structurally valuable under skill shortages and demographic pressure on the manufacturing workforce \cite{yang2025mfgverse,solomashenko2025assisted}. Acceptance studies show that workers with higher psychological capital and clearer perceived usefulness adopt AR more readily and report higher performance gains \cite{khafaga2024psycap,marto2023aram,quandt2021useracceptance}.

\textbf{Dangers.}
Total cost of ownership is dominated by integration, content authoring, change management and lifecycle maintenance, not device price; commercial application quality and usability are cited as the dominant adoption blockers \cite{roberto2024commercialar,machac2025arampaper}. Many pilots fail to graduate because they are evaluated as IT projects rather than sociotechnical system changes \cite{okeeffe2024arhmdshipbuilding,quandt2021useracceptance}.

\textbf{Design considerations.}
Specify the operating model before the device: who authors and maintains content, who certifies overlays, who responds to incidents. Treat AR rollout as sociotechnical change, with explicit user-acceptance Key Performance Indicators (KPIs) derived from Unified Theory of Acceptance and Use of Technology (UTAUT)/Technology Acceptance Model (TAM)-style constructs \cite{quandt2021useracceptance,marto2023aram}. Budget for content authoring tools and version control as first-class infrastructure, on par with the headset fleet \cite{roberto2024commercialar}.

\subsection{A5: Data, Security and Privacy}
\label{sec:a5data}

\textbf{Benefits.}
Coupling AR with digital twins and edge analytics enables fine-grained, context-aware decision support, predictive-maintenance triggers and traceable quality records \cite{emporio2024xrfactory,xu2024arcam,kour2025metaverseindustrial}.

\textbf{Dangers.}
Extended Reality devices are sensor platforms. Eye-tracking alone has been shown to enable activity inference and application fingerprinting \cite{peng2025peekxr,kundu2025privatexr}. Broader systematisations document threats spanning device, network, application and cloud layers, including biometric inference, side-channel leakage and identity re-identification \cite{anjum2025xrhealthsok,elhajj2024xrsec,noah2024biometricxr}. In industrial contexts these risks compound with worker-monitoring concerns and data-residency requirements of regulated sectors \cite{nleya2024industrialmetaverse,awadallah2025metaversecyber}.

\textbf{Design considerations.}
Adopt data minimisation by default: collect only signals required for the immediate task and prefer on-device or edge inference over cloud streaming where feasible \cite{kundu2025privatexr,warin2024privxr}. Apply differential privacy or Explainable Artificial Intelligence-guided mechanisms to sensitive biometric streams \cite{kundu2025privatexr}. Provide user-visible privacy panels and treat eye-tracking, hand-tracking and ambient audio/video as separately consented sensors \cite{warin2024privxr,peng2025peekxr}. Specify cryptographic baselines, authentication and network-layer anomaly detection \cite{elhajj2024xrsec,awadallah2025metaversecyber}.

\subsection{A6: Governance, Ethics and Long-Term Risk}
\label{sec:a6gov}

\textbf{Benefits.}
A coherent governance layer enables scaling: incident reporting, overlay certification, change-management procedures and auditability are the mechanisms by which AR pilots become dependable infrastructure rather than localised demos \cite{kour2025metaverseindustrial,nleya2024industrialmetaverse}.

\textbf{Dangers.}
Industrial-metaverse infrastructure has a non-trivial and contested environmental footprint, driven by data-centre energy use and device manufacturing; without deliberate sustainability practices, the energy demands of supporting Artificial Intelligence/Extended Reality stacks could offset operational efficiency gains \cite{nleya2024industrialmetaverse,guo2024industrialmetaverse}. Worker surveillance, algorithmic management and labour displacement are open governance issues that the literature has so far treated lightly \cite{nleya2024industrialmetaverse,kour2025metaverseindustrial}. Standardised risk-evaluation frameworks for Extended Reality remain immature: a recent XR-in-healthcare systematisation (with findings plausibly transferable to industrial Extended Reality given shared device stacks) found that over 70\% of proposed countermeasures lack standardised evaluations, and that reproducibility is hampered by scarce artefact releases \cite{anjum2025xrhealthsok}..

\textbf{Design considerations.}
Anchor governance in existing standards where they apply (e.g.\ ISO~45001-style Occupational Health and Safety management for Head-Mounted Display use, ISO~9241-style usability and ergonomics for visual displays \cite{antonov2017iso45001,mugeere2023iso9241}) and extend them with AR-specific overlays for content certification, incident reporting and exposure tracking. Make sustainability an explicit axis of vendor selection (renewable energy, device lifecycle, e-waste) \cite{nleya2024industrialmetaverse}. Treat worker monitoring as a co-determination issue, not a unilateral IT decision.

% =============================================================================
\section{Cross-Axis Interactions and Deployment Checklist}
\label{sec:checklist}

The six axes are not independent in operation. Three recurring couplings deserve explicit attention:

\begin{itemize}[leftmargin=*]
  \item \textbf{Tracking accuracy $\leftrightarrow$ attentional safety.} Tracking drift in large dynamic environments \cite{feigl2020localizationlimits} couples directly to attentional tunnelling \cite{wang2021ibarhud,faria2025ibhuds}: an overlay drifting off its referent can mask, rather than highlight, a safety-critical object. Latency and registration accuracy are therefore safety variables, not only UX variables.
  \item \textbf{Edge offloading $\leftrightarrow$ data minimisation.} Offloading rendering and inference to multi-access edge computing reduces device load and battery cost \cite{sampath2024splitxr,trinh2023drlxr} but expands the network attack surface and data-residency footprint \cite{kundu2025privatexr,anjum2025xrhealthsok}. Decisions on where to compute must be made jointly with decisions on what to collect.
  \item \textbf{Acceptance $\leftrightarrow$ governance.} Worker acceptance depends on perceived usefulness, ease of use and trust \cite{quandt2021useracceptance,marto2023aram}; trust depends on visible governance over monitoring, consent and incident handling \cite{nleya2024industrialmetaverse,warin2024privxr}. Acceptance and governance are therefore complementary investments.
\end{itemize}

Table~\ref{tab:checklist} consolidates the framework into a deployment checklist with concrete questions and evidence-grounded acceptance criteria. The checklist is intended as a \emph{conditional menu} for scoping and vendor selection, not a mandatory gate: rows apply as a function of deployment class. A single-station maintenance overlay does not require full 5G slicing or Human-Robot Collaboration (HRC) speed-and-separation analysis; a shop-floor Human-Robot Collaboration deployment does. The mapping from deployment class to applicable rows is left to the practitioner; what the framework asserts is that an axis (A1--A6) cannot be silently skipped without an explicit, defensible rationale.

\begin{table*}[!t]
\centering
\caption{Industrial AR deployment checklist (excerpt). Each row encodes a question, an evidence-grounded acceptance criterion and key references. Rows apply conditionally on deployment class; axes (A1--A6) cannot be silently skipped.}
\label{tab:checklist}
\renewcommand{\arraystretch}{1.15}
\begin{tabular}{p{0.08\linewidth} p{0.32\linewidth} p{0.40\linewidth} p{0.12\linewidth}}
\toprule
\textbf{Axis} & \textbf{Scoping question} & \textbf{Acceptance criterion / signal} & \textbf{Key refs} \\
\midrule
A1 & What measurable operational outcome are we targeting? & A KPI baseline exists; AR pilot reports task time, error, NASA-Task Load Index, retention (not only satisfaction). & \cite{alessa2023arcogload,howard2023metaanalysis} \\
A2 & Is tracking validated in the actual environment? & Pose error budget specified for the cell scale (cf.\ scaling errors of order $10^{-1}$~m/m and position errors of order $10$~m at $\sim 10^3$~m\textsuperscript{2} reported for off-the-shelf Simultaneous Localization and Mapping (SLAM)); tested in production lighting, scale and motion. & \cite{feigl2020localizationlimits,haischt2023automotivetracking} \\
A2 & Is the network budget realistic? & Motion-to-photon and end-to-end latency measured under plant radio conditions. & \cite{sampath2024splitxr,trinh2023drlxr} \\
A2 & Is the stack interoperable? & OpenXR support; documented APIs to Manufacturing Execution System/digital twin; no proprietary lock-in for content. & \cite{jang2023crossplatformxr,emporio2024xrfactory} \\
A3 & Is cognitive load designed, not assumed? & Overlay design follows Cognitive Load Theory; multi-modal load measurement in evaluation. & \cite{alessa2023arcogload,candido2025multimedialearning} \\
A3 & Is attentional tunnelling treated as a hazard? & Hazard detection tested in dual-task conditions; peripheral salience preserved. & \cite{wang2021ibarhud,faria2025ibhuds} \\
A3 & Are cobot safety zones dynamic and standards-aligned? & MR safety zones tied to speed-and-separation monitoring. & \cite{cogurcu2023safetyzones,orsolits2025dynamicsafetyzones} \\
A3 & Is cybersickness exposure managed? & Exposure schedule; Simulator Sickness Questionnaire (SSQ) monitoring; mitigation strategies documented. & \cite{cossio2025cybersickness,kirollos2025arhmdcybersickatsea} \\
A4 & Is content authoring funded as infrastructure? & Authoring tools, certification, versioning resourced; not ad hoc. & \cite{roberto2024commercialar,machac2025arampaper} \\
A4 & Are acceptance KPIs explicit? & Unified Theory of Acceptance and Use of Technology (UTAUT)/Technology Acceptance Model (TAM)-derived measures collected pre/post rollout. & \cite{quandt2021useracceptance,marto2023aram} \\
A5 & Is data minimisation the default? & Eye/hand/audio consented separately; on-device inference preferred. & \cite{kundu2025privatexr,warin2024privxr} \\
A5 & Are Extended Reality-specific threats modelled? & Threat model covers biometric inference and side-channels. & \cite{anjum2025xrhealthsok,elhajj2024xrsec} \\
A6 & Are existing Occupational Health and Safety (OHS)/ergonomics standards mapped? & ISO~45001/9241-style controls extended for Head-Mounted Display use. & \cite{antonov2017iso45001,mugeere2023iso9241} \\
A6 & Is sustainability a vendor-selection criterion? & Energy, lifecycle, e-waste reported. & \cite{nleya2024industrialmetaverse} \\
\bottomrule
\end{tabular}
\end{table*}

% =============================================================================
\section{Open Problems and Research Agenda}
\label{sec:open}

The synthesis surfaces several gaps where the evidence base is too thin to support confident deployment decisions.

\textbf{Longitudinal effects.} Almost all benefit estimates derive from short-term lab studies; longitudinal data on skill retention, productivity persistence and chronic ergonomic effects of daily Head-Mounted Display use are scarce \cite{howard2023metaanalysis,fang2023hmdreview}.

\textbf{Attentional safety standards.} There is no widely adopted, evidence-based standard for evaluating attentional tunnelling and inattentional blindness in industrial AR analogous to those evolving for automotive AR-HUDs \cite{wang2021ibarhud,faria2025ibhuds,ye2025arhudsafety}. Industrial settings, with cobots and moving machinery, plausibly need stricter, not weaker, criteria.

\textbf{Tracking in dynamic factories.} The factory-scale localisation gap reported in \cite{feigl2020localizationlimits} (metre-level position errors and percent-level scaling drift for general-purpose Simultaneous Localization and Mapping (SLAM) in $\sim 10^3$~m\textsuperscript{2} environments) has not been closed by subsequent commercial stacks. Industry-grade benchmarks, hybrid tracking modalities (including reliable bare-hand input \cite{wachter2024spatialar}) and uncertainty-aware overlays remain open \cite{haischt2023automotivetracking}.

\textbf{Extended Reality privacy benchmarks.} Extended Reality privacy lacks standardised risk-evaluation methodologies; reproducibility is limited by scarce artefact releases \cite{anjum2025xrhealthsok,kundu2025privatexr}. Industrial deployments urgently need benchmark suites for biometric-inference risk under realistic enterprise threat models.

\textbf{Industrial-metaverse governance and sustainability.} The environmental and labour-governance implications of metaverse-scale industrial deployments remain under-studied; the current literature is largely descriptive rather than prescriptive \cite{nleya2024industrialmetaverse,kour2025metaverseindustrial}.

\textbf{Contrarian view: is Head-Mounted Display-based AR the right interface at all?}
Spatial AR (projector-based), assisted-reality smart glasses and laser-projection guidance remain competitive on several industrial criteria, ergonomics, hygiene, group visibility, cost. Direct head-to-head comparisons report that spatial/projector-based AR can outperform handheld and Head-Mounted Display alternatives on assembly time, error rate and NASA-Task Load Index, with higher user preference \cite{alves2019comparing,alves2019sparvshandheld,funk2016interactive,solomashenko2025assisted}. Evidence on laser-projection guidance is narrower: in a VR-emulated task it outperformed video and pictorial instructions on cognitive load only in the low-complexity condition, with no significant advantage under high cognitive load \cite{sahoo2025laser}. The assumption that the industrial future is head-mounted should therefore be treated as a hypothesis, not a premise.

% =============================================================================
\section{Conclusion}
\label{sec:conclusion}

Industrial AR is past the proof-of-concept stage but short of dependable infrastructure. The bottleneck is no longer whether AR can outperform paper in isolated assembly tasks; it is whether a deployment integrates technical, human, organisational, security and governance considerations coherently enough to survive contact with a real factory. The full-stack framework proposed here (six distinct but coupled axes, each addressed under benefits, dangers and design considerations, and a deployment checklist with evidence-grounded acceptance criteria) is intended as a practical tool for engineering managers and vendors and as a research scaffold for the open problems identified. The near-term opportunity is not to add more application demos, but to close the empirical and standards gaps that currently force every serious deployment to relitigate the same risks in isolation.

% =============================================================================
\section*{Acknowledgements}
The authors thank colleagues at MixedForm LTD in particular M.H., for their support and valuable discussions and input. No external funding was received for this work.
\subsection{Author Contributions}
\label{subsec:AC}
N.C. led the conception of the framework, literature synthesis, and drafting of the manuscript. M.P. contributed to the industrial context analysis, deployment checklist design, and critical revision of the manuscript. Both authors read and approved the final version.

\subsection{Use of AI-Assisted Tools}
\label{subsec:AIuse}
During the preparation of this work the authors used Perplexity AI, Anthropic Claude and OpenAI ChatGPT for the following purposes: literature search and discovery of relevant references, verification of citation details and factual claims, and assistance with drafting and editing portions of the manuscript. After using these tools, the authors reviewed and edited the content as necessary and take full responsibility for the accuracy, integrity and originality of the published work.

% =============================================================================

\bibliographystyle{IEEEtran}
\bibliography{refs_zotero}

@article{azuma1997survey,
  author  = {Azuma, Ronald T.},
  title   = {A Survey of Augmented Reality},
  journal = {Presence: Teleoperators and Virtual Environments},
  volume  = {6},
  number  = {4},
  pages   = {355--385},
  year    = {1997},
  doi     = {10.1162/pres.1997.6.4.355}
}

@inproceedings{milgram1994taxonomy,
  author    = {Milgram, Paul and Kishino, Fumio},
  title     = {A Taxonomy of Mixed Reality Visual Displays},
  booktitle = {IEICE Transactions on Information and Systems},
  volume    = {E77-D},
  number    = {12},
  pages     = {1321--1329},
  year      = {1994}
}

@article{egger2020ar,
  author  = {Egger, John W. and Masood, Tariq},
  title   = {Augmented reality in support of intelligent manufacturing -- A systematic literature review},
  journal = {Computers \& Industrial Engineering},
  volume  = {140},
  pages   = {106195},
  year    = {2020},
  doi     = {10.1016/j.cie.2019.106195}
}

@article{baroroh2020smartmanuf,
  author  = {Baroroh, Dawi K. and Chu, Chih-Hsing and Wang, Lihui},
  title   = {Systematic literature review on augmented reality in smart manufacturing: Collaboration between human and computational intelligence},
  journal = {Journal of Manufacturing Systems},
  volume  = {61},
  pages   = {696--711},
  year    = {2020},
  doi     = {10.1016/j.jmsy.2020.10.017}
}

@article{fang2023hmdreview,
  author  = {Fang, Wei and Chen, Lixi and Zhang, Tienong and Chen, Cheng and Teng, Zhan and Wang, Lihui},
  title   = {Head-mounted display augmented reality in manufacturing: A systematic review},
  journal = {Robotics and Computer-Integrated Manufacturing},
  volume  = {83},
  pages   = {102567},
  year    = {2023},
  doi     = {10.1016/j.rcim.2023.102567}
}

@article{ho2022quality,
  author  = {Ho, Phuong Thao and Albajez, Jos{\'e} A. and Santolaria, Jorge and Yag{\"u}e-Fabra, Jos{\'e} A.},
  title   = {Study of Augmented Reality Based Manufacturing for Further Integration of Quality Control 4.0: A Systematic Literature Review},
  journal = {Applied Sciences},
  volume  = {12},
  number  = {4},
  pages   = {1961},
  year    = {2022},
  doi     = {10.3390/app12041961}
}

@article{runji2022maintenance,
  author  = {Runji, Joseph and Lee, Yun-Ju and Chu, Chih-Hsing},
  title   = {Systematic Literature Review on Augmented Reality-Based Maintenance Applications in Manufacturing Centered on Operator Needs},
  journal = {International Journal of Precision Engineering and Manufacturing-Green Technology},
  volume  = {10},
  pages   = {567--585},
  year    = {2022},
  doi     = {10.1007/s40684-022-00444-w}
}

@inproceedings{buettner2022vrarmaint,
  author    = {Buettner, Ricardo and Breitenbach, Johannes and Wannenwetsch, Kai and Ostermann, Isabel and Priel, Rene},
  title     = {A Systematic Literature Review of Virtual and Augmented Reality Applications for Maintenance in Manufacturing},
  booktitle = {IEEE Annual Computers, Software, and Applications Conference (COMPSAC)},
  pages     = {686--695},
  year      = {2022},
  doi       = {10.1109/COMPSAC54236.2022.00099}
}

@article{solomashenko2025assisted,
  author  = {Solomashenko, Artem and Afanaseva, Olga and Shishova, Maria and Gulianskii, Igor E.},
  title   = {A systematic review of assisted reality applications in manufacturing},
  journal = {Virtual Reality},
  year    = {2025},
  doi     = {10.1007/s10055-025-01281-3}
}

@article{quandt2021useracceptance,
  author  = {Quandt, Moritz and Freitag, Michael},
  title   = {A Systematic Review of User Acceptance in Industrial Augmented Reality},
  journal = {Frontiers in Education},
  volume  = {6},
  pages   = {700760},
  year    = {2021},
  doi     = {10.3389/feduc.2021.700760}
}

@article{guo2024industrialmetaverse,
  author  = {Guo, Ju-e and Leng, Jiewu and Zhao, J. Leon and Zhou, Xueliang and Yuan, Yu and Lu, Yuqian and Mourtzis, Dimitris and Qi, Qinglin and Huang, Sihan and Song, Xueguan and Liu, Qiang and Wang, Lihui},
  title   = {Industrial metaverse towards {I}ndustry 5.0: Connotation, architecture, enablers, and challenges},
  journal = {Journal of Manufacturing Systems},
  volume  = {76},
  pages   = {69--96},
  year    = {2024},
  doi     = {10.1016/j.jmsy.2024.07.007}
}

@article{nleya2024industrialmetaverse,
  author  = {Nleya, Sindiso M. and Velempini, Mthulisi},
  title   = {Industrial Metaverse: A Comprehensive Review, Environmental Impact, and Challenges},
  journal = {Applied Sciences},
  volume  = {14},
  number  = {13},
  pages   = {5736},
  year    = {2024},
  doi     = {10.3390/app14135736}
}

@article{kour2025metaverseindustrial,
  author  = {Kour, Ravdeep and Karim, Ramin and Venkatesh, N. and Kumar, Uday},
  title   = {Metaverse in industrial contexts -- a comprehensive review},
  journal = {Frontiers in Virtual Reality},
  volume  = {6},
  pages   = {1488926},
  year    = {2025},
  doi     = {10.3389/frvir.2025.1488926}
}

@article{yang2025mfgverse,
  author  = {Yang, Hui and Aqlan, Faisal and Zhao, Richard},
  title   = {Towards Smart Manufacturing Metaverse via Digital Twinning in Extended Reality},
  journal = {Journal of Computing and Information Science in Engineering},
  volume  = {25},
  number  = {12},
  pages   = {120813},
  year    = {2025},
  doi     = {10.1115/1.4070437}
}

@article{yang2019stages,
  author  = {Yang, Zhen and Shi, Jinlei and Jiang, Wenjun and Sui, Yuexin and Wu, Yimin and Ma, Shu and Kang, Chunyan and Li, Hongting},
  title   = {Influences of Augmented Reality Assistance on Performance and Cognitive Loads in Different Stages of Assembly Task},
  journal = {Frontiers in Psychology},
  volume  = {10},
  pages   = {1703},
  year    = {2019},
  doi     = {10.3389/fpsyg.2019.01703}
}

@article{alessa2023arcogload,
  author  = {Alessa, Faisal M. and Alhaag, Mohammed H. and Al-harkan, Ibrahim M. and Ramadan, Mohamed and Alqahtani, Fahad M.},
  title   = {A Neurophysiological Evaluation of Cognitive Load during Augmented Reality Interactions in Various Industrial Maintenance and Assembly Tasks},
  journal = {Sensors},
  volume  = {23},
  number  = {18},
  pages   = {7698},
  year    = {2023},
  doi     = {10.3390/s23187698}
}

@article{sahoo2025laser,
  author  = {Sahoo, Praneet and Biondi, Francesco N.},
  title   = {On the effect of using an augmented reality laser projection operator guidance system on cognitive workload and assembly task performance},
  journal = {Applied Ergonomics},
  volume  = {130},
  pages   = {104662},
  year    = {2025},
  doi     = {10.1016/j.apergo.2025.104662}
}

@article{seeliger2023qualityinspection,
  author  = {Seeliger, Anika and Cheng, Long and Netland, Torbj{\o}rn H.},
  title   = {Augmented reality for industrial quality inspection: An experiment assessing task performance and human factors},
  journal = {Computers in Industry},
  volume  = {151},
  pages   = {103985},
  year    = {2023},
  doi     = {10.1016/j.compind.2023.103985}
}

@article{okeeffe2024arhmdshipbuilding,
  author  = {O'Keeffe, Valerie and Jang, Ryan and Manning, Kosta and Trott, Robert and Howard, Sara and Hordacre, Ann-Louise and Spoehr, John},
  title   = {Forming a view: a human factors case study of augmented reality collaboration in assembly},
  journal = {Ergonomics},
  year    = {2024},
  doi     = {10.1080/00140139.2024.2352733}
}

@article{howard2023metaanalysis,
  author  = {Howard, Matt C. and Davis, Maggie M.},
  title   = {A Meta-analysis of augmented reality programs for education and training},
  journal = {Virtual Reality},
  year    = {2023},
  doi     = {10.1007/s10055-023-00844-6}
}

@article{baashar2022armedical,
  author  = {Baashar, Yahia and Alkawsi, Gamal and Wan Ahmad, Wan Noorina and Alhussian, Hitham and Alwadain, Ayed and Capretz, Luiz Fernando and Babiker, Ammar and Alghail, Adnan},
  title   = {Effectiveness of Using Augmented Reality for Training in the Medical Professions: Meta-analysis},
  journal = {JMIR Serious Games},
  volume  = {10},
  number  = {3},
  pages   = {e32715},
  year    = {2022},
  doi     = {10.2196/32715}
}

@article{gong2024arsafetytraining,
  author  = {Gong, Peizhen and Lu, Ying and Lovreglio, Ruggiero and Lv, Xiaofeng and Chi, Zexu},
  title   = {Applications and effectiveness of augmented reality in safety training: A systematic literature review and meta-analysis},
  journal = {Safety Science},
  volume  = {178},
  pages   = {106624},
  year    = {2024},
  doi     = {10.1016/j.ssci.2024.106624}
}

@article{candido2025multimedialearning,
  author  = {Candido, Vito and Cattaneo, Alberto},
  title   = {Applying cognitive theory of multimedia learning principles to augmented reality and its effects on cognitive load and learning outcomes},
  journal = {Computers in Human Behavior Reports},
  volume  = {18},
  pages   = {100678},
  year    = {2025},
  doi     = {10.1016/j.chbr.2025.100678}
}

@inproceedings{feigl2020localizationlimits,
  author    = {Feigl, Tobias and Porada, Andreas and Steiner, Steve and Loeffler, Christoffer and Mutschler, Christopher and Philippsen, Michael},
  title     = {Localization Limitations of {ARC}ore, {ARK}it, and {H}ololens in Dynamic Large-scale Industry Environments},
  booktitle = {Proc. 15th Intl. Joint Conf. Computer Vision, Imaging and Computer Graphics Theory and Applications (VISIGRAPP)},
  year      = {2020},
  doi       = {10.5220/0008989903070318}
}

@inproceedings{haischt2023automotivetracking,
  title={What’s (Not) Tracking? Factors of Influence in Industrial Augmented Reality Tracking: A Use Case Study in an Automotive Environment},
  author={Haischt, Jonas and Sedlmair, Michael},
  booktitle={Proceedings of the 15th International Conference on Automotive User Interfaces and Interactive Vehicular Applications},
  pages={42--51},
  year={2023},
  doi       = {10.1145/3580585.3607156}
}

@inproceedings{wachter2024spatialar,
  author    = {W{\"a}chter, Christoph and Deppe, Sahar and R{\"o}cker, Carsten},
  title     = {Spatial Augmented Reality in Industrial Environments -- An Approach for Improved Hand-Tracking Using Combined Computer Vision Technologies},
  booktitle = {IEEE VR Workshops},
  year      = {2024},
  doi       = {10.1109/VRW62533.2024.00045}
}

@article{sampath2024splitxr,
  author  = {Sampath, Hemanth and Tinnakornsrisuphap, Peerapol and Hande, Prashanth},
  title   = {Enabling Extended Reality Over {5G} with Distributed Computing},
  journal = {IEEE Communications Magazine},
  year    = {2024},
  doi     = {10.1109/MCOM.003.2300747}
}

@article{trinh2023drlxr,
  author  = {Trinh, Bao and Muntean, Gabriel-Miro},
  title   = {A Deep Reinforcement Learning-Based Offloading Scheme for Multi-Access Edge Computing-Supported eXtended Reality Systems},
  journal = {IEEE Transactions on Vehicular Technology},
  year    = {2023},
  doi     = {10.1109/TVT.2022.3207692}
}

@inproceedings{aktas2025aislicingxr,
  author    = {Aktas, Semih and Kosu, Semiha and Kalem, G{\"o}khan},
  title     = {Experimental Evaluation of {AI}-Based Network Slicing for an {XR} Platform Leveraging Beyond {5G} and Edge Computing},
  booktitle = {IEEE ComManTel},
  year      = {2025},
  doi       = {10.1109/ComManTel68363.2025.11368512}
}

@article{jang2023crossplatformxr,
  author  = {Jang, Heejin and Lee, Sunghee and Shin, Jung-Hun and Lee, Seungmin and Kim, Min-Ah and Nam, Dukyun},
  title   = {Development of Cross-platform {XR} Content Supporting Heterogeneous Devices},
  journal = {Journal of Digital Contents Society},
  volume  = {24},
  number  = {10},
  pages   = {2533--2541},
  year    = {2023},
  doi     = {10.9728/dcs.2023.10.24.10.2533}
}

@article{koo2025iiotinterop,
  author  = {Koo, Jahoon and Kang, Giluk and Kim, Young-Gab},
  title   = {Access Control Framework for Cross-Platform Interoperability in the Industrial {I}nternet of {T}hings},
  journal = {IEEE Transactions on Industrial Informatics},
  year    = {2025},
  doi     = {10.1109/TII.2024.3461783}
}

@article{emporio2024xrfactory,
  author  = {Emporio, Marco and Caputo, Ariel and Pintani, Deborah and Cheng, Dong Seon and De Marchi, Thomas and Forte, Gianmaria and Fummi, Franco and Giachetti, Andrea},
  title   = {Integration of Extended Reality with a Cyber-Physical Factory Environment and its Digital Twins},
  journal = {ACM Transactions on Multimedia Computing, Communications, and Applications},
  year    = {2024},
  doi     = {10.1145/3660246}
}

@article{xu2024arcam,
  author  = {Xu, Shengyang and Lu, Yuan and Yu, Chunyang},
  title   = {Augmented reality-assisted cloud additive manufacturing with digital twin technology for multi-stakeholder value co-creation in product innovation},
  journal = {Heliyon},
  volume  = {10},
  number  = {3},
  pages   = {e25722},
  year    = {2024},
  doi     = {10.1016/j.heliyon.2024.e25722}
}

@article{cossio2025cybersickness,
  author  = {Cossio, Samantha and Chiappinotto, Stefania and Dentice, Sara and Moreal, Chiara and Magro, Gaia and Dussi, Gaia and Palese, Alvisa and Galazzi, Alessandro},
  title   = {Cybersickness and discomfort from head-mounted displays delivering fully immersive virtual reality: A systematic review},
  journal = {Nurse Education in Practice},
  volume  = {85},
  pages   = {104376},
  year    = {2025},
  doi     = {10.1016/j.nepr.2025.104376}
}

@article{kirollos2025arhmdcybersickatsea,
  author  = {Kirollos, Ramy and Merchant, Wasim},
  title   = {Augmented reality head-mounted display at-sea use causes cybersickness},
  journal = {Applied Ergonomics},
  volume  = {125},
  pages   = {104484},
  year    = {2025},
  doi     = {10.1016/j.apergo.2025.104484}
}

@article{wang2021ibarhud,
  author  = {Wang, Yuwei and Wu, Yimin and Chen, Cheng and Wu, Bohan and Ma, Shu and Wang, Duming and Li, Hongting and Yang, Zhen},
  title   = {Inattentional Blindness in Augmented Reality Head-Up Display-Assisted Driving},
  journal = {International Journal of Human--Computer Interaction},
  year    = {2021},
  doi     = {10.1080/10447318.2021.1970434}
}

@article{chen2023ibposition,
  author  = {Chen, Wanting and Song, Jiaqing and Wang, Yuwei and Wu, Changxu and Ma, Shu and Wang, Duming and Yang, Zhen and Li, Hongting},
  title   = {Inattentional blindness to unexpected hazard in augmented reality head-up display assisted driving: The impact of the relative position between stimulus and augmented graph},
  journal = {Traffic Injury Prevention},
  year    = {2023},
  doi     = {10.1080/15389588.2023.2186735}
}

@article{faria2025ibhuds,
  author  = {de Oliveira Faria, Nayara and Gabbard, Joseph L.},
  title   = {Inattentional Blindness with Augmented Reality {HUDS}: An On-road Study},
  journal = {arXiv preprint arXiv:2505.00879},
  year    = {2025},
  doi     = {10.48550/arXiv.2505.00879}
}

@article{ye2025arhudsafety,
  author  = {Ye, Menlong and Yin, Jun},
  title   = {Spatial Plane Positioning of {AR}-{HUD} Graphics: Implications for Driver Inattentional Blindness in Navigation and Collision Warning Scenarios},
  journal = {Electronics},
  volume  = {14},
  number  = {23},
  pages   = {4768},
  year    = {2025},
  doi     = {10.3390/electronics14234768}
}

@inproceedings{cogurcu2023safetyzones,
  title={Augmented reality safety zone configurations in human-robot collaboration: A user study},
  author={Cogurcu, Yunus Emre and Maddock, Steve},
  booktitle={Companion of the 2023 ACM/IEEE International Conference on Human-Robot Interaction},
  pages={360--363},
  year={2023},
  doi     = {10.1145/3568294.3580106}
}

@inproceedings{orsolits2025dynamicsafetyzones,
  author    = {Orsolits, Horst and Saliger, Alexandra and Korn, Alexander},
  title     = {Interactive Mixed Reality Visualization of Dynamic Safety Zones in Human-Robot Collaboration},
  booktitle = {IEEE ISMAR Adjunct},
  year      = {2025},
  doi       = {10.1109/ISMAR-Adjunct68609.2025.00090}
}

@article{zhang2025mrhrc,
  author  = {Zhang, Kaiyuan and Yan, Yuchen and Jia, Yunyi},
  title   = {Mixed-Reality ({MR}) Enhanced Human--Robot Collaboration: Communicating Robot Intentions to Humans},
  journal = {Robotics},
  volume  = {14},
  number  = {10},
  pages   = {133},
  year    = {2025},
  doi     = {10.3390/robotics14100133}
}

@inproceedings{anjum2025xrhealthsok,
  title={Beyond the Headset: A Systematization of Knowledge on Extended Reality Privacy and Security in Healthcare},
  author={Anjum, Nafisa and Mahmud, M Rasel},
  booktitle={Proceedings of the 2025 31st ACM Symposium on Virtual Reality Software and Technology},
  pages={1--12},
  year={2025},
  doi     = {10.1145/3756884.3766045}
}

@article{elhajj2024xrsec,
  author  = {El-hajj, Mohammed},
  title   = {Cybersecurity and Privacy Challenges in Extended Reality: Threats, Solutions, and Risk Mitigation Strategies},
  journal = {Virtual Worlds},
  volume  = {4},
  number  = {1},
  pages   = {1},
  year    = {2024},
  doi     = {10.3390/virtualworlds4010001}
}

@article{kundu2025privatexr,
  author  = {Kundu, Ripan Kumar and Ahmed, Istiak and Hoque, Khaza Anuarul},
  title   = {{P}rivate{XR}: Defending Privacy Attacks in Extended Reality Through Explainable {AI}-Guided Differential Privacy},
  journal = {IEEE ISMAR},
  year    = {2025},
  doi     = {10.1109/ISMAR67309.2025.00061}
}

@inproceedings{peng2025peekxr,
  author    = {Peng, Chuyang and Ali, Mutahar and Farrukh, Habiba},
  title     = {Poster: {P}eek{XR}: Understanding Privacy Leakages from Eye Gaze in Extended Reality},
  booktitle = {Proc. ACM MobiSys},
  year      = {2025},
  doi       = {10.1145/3711875.3734567}
}

@article{noah2024biometricxr,
  author  = {Noah, Naheem and Das, Sanchari},
  title   = {Privacy and Security in Extended Reality: Exploring the Risks of External Biometric Data Collection},
  journal = {SSRN preprint},
  year    = {2024},
  doi     = {10.2139/ssrn.4780358}
}

@inproceedings{warin2024privxr,
  author    = {Warin, Chris and Seeger, David and Shams, Shirin and Reinhardt, Delphine},
  title     = {{P}riv{XR}: A Cross-Platform Privacy-Preserving {API} and Privacy Panel for Extended Reality},
  booktitle = {IEEE PerCom Workshops},
  year      = {2024},
  doi       = {10.1109/PerComWorkshops59983.2024.10503341}
}

@article{awadallah2025metaversecyber,
  author  = {Awadallah, Abeer and Eledlebi, Khouloud and Zemerly, Mohammed Jamal and Puthal, Deepak and Damiani, Ernesto and Taha, Kamal and Kim, Tae-Yeon and Yoo, Paul D. and Choo, Kim-Kwang Raymond and Yim, Man-Sung and Yeun, Chan Yeob},
  title   = {Artificial Intelligence-Based Cybersecurity for the Metaverse: Research Challenges and Opportunities},
  journal = {IEEE Communications Surveys \& Tutorials},
  year    = {2025},
  doi     = {10.1109/COMST.2024.3442475}
}

@article{marto2023aram,
  author  = {Marto, Anabela and Gon{\c c}alves, Alexandrino and Melo, Miguel and Bessa, Maximino and Silva, Rui J. R.},
  title   = {{ARAM}: A Technology Acceptance Model to Ascertain the Behavioural Intention to Use Augmented Reality},
  journal = {Journal of Imaging},
  volume  = {9},
  number  = {3},
  pages   = {73},
  year    = {2023},
  doi     = {10.3390/jimaging9030073}
}

@inproceedings{khafaga2024psycap,
  author    = {Khafaga, Ahmed A. A. E. M. A. and Horan, Ben and Caires Moreira, Lorena and Kauffman, Marcos and Nienaber, Ann-Marie and Mortimer, Michael},
  title     = {Unlocking Human Potential: Psychological Capital's Influence on Augmented Reality Adoption in Manufacturing Environments},
  booktitle = {IEEE ISMAR Adjunct},
  year      = {2024},
  doi       = {10.1109/ISMAR-Adjunct64951.2024.00128}
}

@article{roberto2024commercialar,
  author  = {Roberto, Rafael and Breyer, Felipe and Betts, Joe and Newnes, Linda and Shokrani, Alborz},
  title   = {A Study of Commercial Augmented Reality Applications in Manufacturing: A Subject Matter Expert Analysis},
  journal = {Procedia CIRP},
  year    = {2024},
  doi     = {10.1016/j.procir.2024.04.003}
}

@article{machac2025arampaper,
  author  = {Mach{\'a}{\v c}, Tom{\'a}{\v s} and Ho{\v r}ej{\v s}{\'\i}, Petr and {\v S}imon, Michal},
  title   = {Augmented Reality-Enabled Visualisation and Process Control in {3D} Printing and Additive Manufacturing},
  journal = {MM Science Journal},
  year    = {2025},
  doi     = {10.17973/mmsj.2025_12_2025090}
}

@article{antonov2017iso45001,
  author  = {Antonov, Anca-Elena and Bejinariu, Costel and Darabont, Doru},
  title   = {Key elements on implementing an occupational health and safety management system using {ISO} 45001 standard},
  journal = {MATEC Web of Conferences},
  volume  = {121},
  pages   = {11007},
  year    = {2017},
  doi     = {10.1051/matecconf/201712111007}
}

@article{mugeere2023iso9241,
  author  = {Mugeere, Anthony and Sch{\"o}nenberger, Klaus and Ortiz-Escobar, Luisa Mar{\'\i}a and Chavarria, Mar{\'\i}a Alejandra and Stein, Michael Ashley and Rivas Velarde, Minerva and Hurst, Samia},
  title   = {Assessing the implementation of user-centred design standards on assistive technology for persons with visual impairments: a systematic review},
  journal = {Frontiers in Rehabilitation Sciences},
  year    = {2023},
  doi     = {10.3389/fresc.2023.1238158}
}

@inproceedings{alves2019sparvshandheld,
  author    = {Alves, Jo{\~a}o B. and Marques, Bernardo and Dias, Paulo and Santos, Beatriz Sousa},
  title     = {Using Augmented Reality for Industrial Quality Assurance: A Shop Floor User Study},
  booktitle = {2019 IEEE International Conference on Autonomous Robot Systems and Competitions (ICARSC)},
  year      = {2019},
  pages     = {1--6},
  publisher = {IEEE},
  doi       = {10.1109/ICARSC.2019.8733642}
}

@article{bailenson2024passthrough,
  author  = {Bailenson, Jeremy N. and Han, Eugy and Boyd, Stephanie and Santoso, Monique and Markowitz, David M. and Beans, Chevy and Queiroz, Anna and Ratan, Rabindra and DeVeaux, Cyan},
  title   = {Seeing the World Through Digital Prisms: Psychological Implications of Passthrough Video Usage in Mixed Reality},
  journal = {Technology, Mind, and Behavior},
  volume  = {5},
  number  = {2},
  year    = {2024},
  doi     = {10.1037/tmb0000129}
}

@inproceedings{funk2016interactive,
  title={Interactive worker assistance: comparing the effects of in-situ projection, head-mounted displays, tablet, and paper instructions},
  author={Funk, Markus and Kosch, Thomas and Schmidt, Albrecht},
  booktitle={Proceedings of the 2016 ACM international joint conference on pervasive and ubiquitous computing},
  pages={934--939},
  year={2016}
}

@inproceedings{el2025mind,
  title={Mind the GAP: Geometry aware passthrough mitigates cybersickness},
  author={El Chemaly, Trishia and Goyal, Mohit and Duan, Tinglin and Phadnis, Vrushank and Khattar, Sakar and Vlaskamp, Bjorn and Kulshrestha, Achin and Turner, Eric Lee and Purohit, Aveek and Neiswander, Gregory and others},
  booktitle={Proceedings of the Extended Abstracts of the CHI Conference on Human Factors in Computing Systems},
  pages={1--11},
  year={2025}
}

@article{szentirmai2025you,
  title={Are you drunk? No, I am CybAR sick!--interacting with the real world via pass-through augmented reality is a sobering discovery},
  author={Szentirmai, Attila Bekkvik and Alsos, Ole Andreas and Torkildsby, Anne Britt and Inal, Yavuz},
  journal={Frontiers in Virtual Reality},
  volume={6},
  pages={1533236},
  year={2025},
  publisher={Frontiers Media SA}
}

@inproceedings{freiwald2018camera,
  title={Camera time warp: compensating latency in video see-through head-mounted-displays for reduced cybersickness effects},
  author={Freiwald, Jann Philipp and Katzakis, Nicholas and Steinicke, Frank},
  booktitle={Proceedings of the 24th ACM symposium on virtual reality software and technology},
  pages={1--7},
  year={2018}
}

@inproceedings{alves2019comparing,
  title={Comparing spatial and mobile augmented reality for guiding assembling procedures with task validation},
  author={Alves, Jo{\~a}o and Marques, Bernardo and Oliveira, Miguel and Ara{\'u}jo, Tiago and Dias, Paulo and Santos, Beatriz Sousa},
  booktitle={2019 IEEE international conference on autonomous robot systems and competitions (ICARSC)},
  pages={1--6},
  year={2019},
  organization={IEEE}
}

\end{document}